\documentclass[a4paper,11pt]{article}

\usepackage[top=1.0in,right=1.0in,left=1.0in,bottom=1.75in]{geometry}
\usepackage{latexsym}
\usepackage{amssymb}
\usepackage{amsmath,amsfonts,amssymb,amsthm}
\usepackage{float}
\usepackage{graphics,graphicx}
\usepackage{tabularx}
\usepackage{amsfonts}
\usepackage{amsmath}
\usepackage{enumerate}
\usepackage{booktabs}
\usepackage{multirow}
\usepackage{sectsty}
\usepackage[affil-it,auth-sc]{authblk}
\usepackage{subcaption}
\usepackage[usenames,dvipsnames]{xcolor}

\usepackage{tabu}

\sectionfont{\normalsize}
\subsectionfont{\normalsize}

\begin{document}

\newcommand{\singlespace}{\baselineskip=12pt\lineskiplimit=0pt\lineskip=0pt}
\def\ds{\displaystyle}

\newcommand{\beq}{\begin{equation}}
\newcommand{\eeq}{\end{equation}}
\newcommand{\lb}{\label}
\newcommand{\ph}{\phantom}
\newcommand{\beqar}{\begin{eqnarray}}
\newcommand{\eeqar}{\end{eqnarray}}
\newcommand{\barr}{\begin{array}}
\newcommand{\earr}{\end{array}}
\newcommand{\jump}{\parallel}
\newcommand{\Ehat}{\hat{E}}
\newcommand{\That}{\hat{\bf T}}
\newcommand{\Ahat}{\hat{A}}
\newcommand{\chat}{\hat{c}}
\newcommand{\shat}{\hat{s}}
\newcommand{\khat}{\hat{k}}
\newcommand{\muhat}{\hat{\mu}}
\newcommand{\mc}{M^{\scriptscriptstyle C}}
\newcommand{\mei}{M^{\scriptscriptstyle M,EI}}
\newcommand{\mec}{M^{\scriptscriptstyle M,EC}}
\newcommand{\hbeta}{{\hat{\beta}}}
\newcommand{\rec}[2]{\left( #1 #2 \ds{\frac{1}{#1}}\right)}
\newcommand{\rep}[2]{\left( {#1}^2 #2 \ds{\frac{1}{{#1}^2}}\right)}
\newcommand{\derp}[2]{\ds{\frac {\partial #1}{\partial #2}}}
\newcommand{\derpn}[3]{\ds{\frac {\partial^{#3}#1}{\partial #2^{#3}}}}
\newcommand{\dert}[2]{\ds{\frac {d #1}{d #2}}}
\newcommand{\dertn}[3]{\ds{\frac {d^{#3} #1}{d #2^{#3}}}}
\newcommand{\ct}{\captionof{table}}
\newcommand{\cf}{\captionof{figure}}
\newcommand{\dd}{\diff}
\newcommand{\rr}{\textcolor{red}}

\def\c{{\circ}}
\def\bob{{\, \underline{\overline{\otimes}} \,}}
\def\ob{{\, \underline{\otimes} \,}}
\def\scalp{\mbox{\boldmath$\, \cdot \, $}}
\def\gdp{\makebox{\raisebox{-.215ex}{$\Box$}\hspace{-.778em}$\times$}}
\def\daa{\makebox{\raisebox{-.050ex}{$-$}\hspace{-.550em}$: ~$}}
\def\mK{\mbox{${\mathcal{K}}$}}
\def\cK{\mbox{${\mathbb {K}}$}}

\def\Xint#1{\mathchoice
   {\XXint\displaystyle\textstyle{#1}}%
   {\XXint\textstyle\scriptstyle{#1}}%
   {\XXint\scriptstyle\scriptscriptstyle{#1}}%
   {\XXint\scriptscriptstyle\scriptscriptstyle{#1}}%
   \!\int}
\def\XXint#1#2#3{{\setbox0=\hbox{$#1{#2#3}{\int}$}
     \vcenter{\hbox{$#2#3$}}\kern-.5\wd0}}
\def\ddashint{\Xint=}
\def\fpint{\Xint=}
\def\dashint{\Xint-}
\def\cpvint{\Xint-}
\def\intl{\int\limits}
\def\cpvintl{\cpvint\limits}
\def\fpintl{\fpint\limits}
\def\ointl{\oint\limits}
\def\bA{{\bf A}}
\def\ba{{\bf a}}
\def\bB{{\bf B}}
\def\bb{{\bf b}}
\def\bc{{\bf c}}
\def\bC{{\bf C}}
\def\bD{{\bf D}}
\def\bE{{\bf E}}
\def\be{{\bf e}}
\def\bbf{{\bf f}}
\def\bF{{\bf F}}
\def\bG{{\bf G}}
\def\bg{{\bf g}}
\def\bi{{\bf i}}
\def\bH{{\bf H}}
\def\bK{{\bf K}}
\def\bL{{\bf L}}
\def\bM{{\bf M}}
\def\bN{{\bf N}}
\def\bn{{\bf n}}
\def\bm{{\bf m}}
\def\b0{{\bf 0}}
\def\bo{{\bf o}}
\def\bX{{\bf X}}
\def\bx{{\bf x}}
\def\bP{{\bf P}}
\def\bp{{\bf p}}
\def\bQ{{\bf Q}}
\def\bq{{\bf q}}
\def\bR{{\bf R}}
\def\bS{{\bf S}}
\def\bs{{\bf s}}
\def\bT{{\bf T}}
\def\bt{{\bf t}}
\def\bU{{\bf U}}
\def\bu{{\bf u}}
\def\bv{{\bf v}}
\def\bw{{\bf w}}
\def\bW{{\bf W}}
\def\by{{\bf y}}
\def\bz{{\bf z}}
\def\T{{\bf T}}
\def\Te{\textrm{T}}
\def\Id{{\bf I}}
\def\bxi{\mbox{\boldmath${\xi}$}}
\def\balpha{\mbox{\boldmath${\alpha}$}}
\def\bbeta{\mbox{\boldmath${\beta}$}}
\def\bepsilon{\mbox{\boldmath${\epsilon}$}}
\def\bvarepsilon{\mbox{\boldmath${\varepsilon}$}}
\def\bomega{\mbox{\boldmath${\omega}$}}
\def\bphi{\mbox{\boldmath${\phi}$}}
\def\bsigma{\mbox{\boldmath${\sigma}$}}
\def\bfeta{\mbox{\boldmath${\eta}$}}
\def\bDelta{\mbox{\boldmath${\Delta}$}}
\def\btau{\mbox{\boldmath $\tau$}}
\def\tr{{\rm tr}}
\def\dev{{\rm dev}}
\def\div{{\rm div}}
\def\Div{{\rm Div}}
\def\Grad{{\rm Grad}}
\def\grad{{\rm grad}}
\def\Lin{{\rm Lin}}
\def\Sym{{\rm Sym}}
\def\Skw{{\rm Skew}}
\def\abs{{\rm abs}}
\def\Re{{\rm Re}}
\def\Im{{\rm Im}}
\def\capB{\mbox{\boldmath${\mathsf B}$}}
\def\capC{\mbox{\boldmath${\mathsf C}$}}
\def\capD{\mbox{\boldmath${\mathsf D}$}}
\def\capE{\mbox{\boldmath${\mathsf E}$}}
\def\capG{\mbox{\boldmath${\mathsf G}$}}
\def\tcapG{\tilde{\capG}}
\def\capH{\mbox{\boldmath${\mathsf H}$}}
\def\capK{\mbox{\boldmath${\mathsf K}$}}
\def\capL{\mbox{\boldmath${\mathsf L}$}}
\def\capM{\mbox{\boldmath${\mathsf M}$}}
\def\capR{\mbox{\boldmath${\mathsf R}$}}
\def\capW{\mbox{\boldmath${\mathsf W}$}}

\def\i{\mbox{${\mathrm i}$}}
\def\mC{\mbox{\boldmath${\mathcal C}$}}
\def\mB{\mbox{${\mathcal B}$}}
\def\mE{\mbox{${\mathcal{E}}$}}
\def\mL{\mbox{${\mathcal{L}}$}}
\def\mK{\mbox{${\mathcal{K}}$}}
\def\mV{\mbox{${\mathcal{V}}$}}
\def\C{\mbox{\boldmath${\mathcal C}$}}
\def\E{\mbox{\boldmath${\mathcal E}$}}

\def\AAM{{\it Advances in Applied Mechanics }}
\def\ACME{{\it Arch. Comput. Meth. Engng.}}
\def\ARMA{{\it Arch. Rat. Mech. Analysis}}
\def\AMR{{\it Appl. Mech. Rev.}}
\def\ASCEEM{{\it ASCE J. Eng. Mech.}}
\def\ACTA{{\it Acta Mater.}}
\def\CMAME {{\it Comput. Meth. Appl. Mech. Engrg.}}
\def\CRAS{{\it C. R. Acad. Sci. Paris}}
\def\CRM{{\it Comptes Rendus M\'ecanique}}
\def\EFM{{\it Eng. Fracture Mechanics}}
\def\EJMA{{\it Eur.~J.~Mechanics-A/Solids}}
\def\IJES{{\it Int. J. Eng. Sci.}}
\def\IJF{{\it Int. J. Fracture}}
\def\IJMS{{\it Int. J. Mech. Sci.}}
\def\IJNAMG{{\it Int. J. Numer. Anal. Meth. Geomech.}}
\def\IJP{{\it Int. J. Plasticity}}
\def\IJSS{{\it Int. J. Solids Structures}}
\def\IngA{{\it Ing. Archiv}}
\def\JAM{{\it J. Appl. Mech.}}
\def\JAP{{\it J. Appl. Phys.}}
\def\JAE{{\it J. Aerospace Eng.}}
\def\JE{{\it J. Elasticity}}
\def\JM{{\it J. de M\'ecanique}}
\def\JMPS{{\it J. Mech. Phys. Solids}}
\def\JSV{{\it J. Sound and Vibration}}
\def\MACRO{{\it Macromolecules}}
\def\MMT{{\it Mech. Mach. Th.}}
\def\MOM{{\it Mech. Materials}}
\def\MMS{{\it Math. Mech. Solids}}
\def\MMT{{\it Metall. Mater. Trans. A}}
\def\MPCPS{{\it Math. Proc. Camb. Phil. Soc.}}
\def\MSE{{\it Mater. Sci. Eng.}}
\def\NATURE{{\it Nature}}
\def\NATUREM{{\it Nature Mater.}}
\def\PHIL{{\it Phil. Trans. R. Soc.}}
\def\PMPS{{\it Proc. Math. Phys. Soc.}}
\def\PNAS{{\it Proc. Nat. Acad. Sci.}}
\def\PRE{{\it Phys. Rev. E}}
\def\PRL{{\it Phys. Rev. Letters}}
\def\PRSL{{\it Proc. R. Soc.}}
\def\RIIT{{\it Rozprawy Inzynierskie - Engineering Transactions}}
\def\ROCK{{\it Rock Mech. and Rock Eng.}}
\def\QAM{{\it Quart. Appl. Math.}}
\def\QJMAM{{\it Quart. J. Mech. Appl. Math.}}
\def\SCIENCE{{\it Science}}
\def\SCRMAT{{\it Scripta Mater.}}
\def\SM{{\it Scripta Metall.}}
\def\ZAMM{{\it Z. Angew. Math. Mech.}}
\def\ZAMP{{\it Z. Angew. Math. Phys.}}
\def\ZVDI{{\it Z. Verein. Deut. Ing.}}

\renewcommand\Affilfont{\itshape\small}
\setlength{\affilsep}{1em}
\renewcommand\Authsep{, }
\renewcommand\Authand{ and }
\renewcommand\Authands{ and }
\setcounter{Maxaffil}{3}

\renewcommand{\vec}[1]{{\boldsymbol{#1}}}

\definecolor{traz}{RGB}{200,50,0}
\definecolor{comp}{RGB}{0,30,200}

\newcommand{\djc}[1]{{\color{Orange}#1}}

\title{Cymatics for the cloaking of flexural vibrations in a structured plate}

\author[1]{D. Misseroni}
\author[2]{D. J. Colquitt}
\author[3]{A. B. Movchan \footnote{Corresponding author. Phone:\,+44\,(0)151\,7944740; E-mail:\,abm@liv.ac.uk; Fax:\,+44\,(0)151\,7944056.}}
\author[3]{\\N. V. Movchan}
\author[4]{I. S. Jones}

\affil[1]{
DICAM, University of Trento,
via~Mesiano~77, I-38123 Trento, Italy.
}
\affil[2]{
Department of Mathematics, Imperial College London, South Kensington, London SW7 2AZ
}
\affil[3]{
Department of Mathematical Sciences, University of Liverpool, Liverpool L69 3BX, UK.
}
\affil[4]{
School of Engineering, Liverpool John Moores University, Liverpool L3 3AF, UK.
}


\date{}
\maketitle

\begin{abstract}

We present a proof of concept design for a structured square cloak enclosing a void in an elastic lattice, subjected to flexural vibrations.
In addition to the theoretical design, we fabricate and experimentally test an elastic invisibility cloak for flexural waves in a mechanical lattice.
Further verification and modelling is performed numerically through finite element simulations.
The primary advantage of our square lattice cloak, over other designs, is the straightforward implementation and the ease of construction.
The elastic lattice cloak shows an impressive efficiency within the predicted frequency range of 100-250Hz.

\end{abstract}


\section{Introduction} \lb{S_INTRO}

Since the first experimental demonstration of the microwave invisibility cloak~\cite{schurig}, there has been an explosion of theoretical and practical advances in the design and analysis of electromagnetic metamaterials~\cite{Fleury & another review article}.
In contrast, the much more challenging problem of creating invisibility cloaks and metamaterials for elastodynamics has been much less studied.
Notable recent advances in the theoretical analysis of cloaks for elastic waves have been made by Milton \emph{et al.}~\cite{MiltonNJP,MiltonPRS}, Norris \& Shuvalov~\cite{NorrisWaveMotion}, Brun \emph{et al.}~\cite{BrunAPL,BrunNJP}, Farhat \emph{et al.}~\cite{FarhatSebastienPlates}, Jones \emph{et al.}~\cite{JonesIntJSolsStruct}, Colquitt \emph{et al.}~\cite{ColquittPRS,ColquittJMPS}, Guennau \emph{et al.}~\cite{Guenneau} and Parnell \emph{et al.}~\cite{ParnellPRS,ParnellWaveMotion,NorrisPRS,NorrisAPL}.
These theoretical developments have been complemented with a series of experimental implementations of multi-scale mechanical cloaks performed by the group led by Wegener~{\cite{Wegener2012,UnfealabilityCloak,WegenerPNAS}}.

\begin{figure}[h]
\renewcommand{\figurename}{\footnotesize{Fig.}}
\begin{center}
\includegraphics[width=10.0 cm]{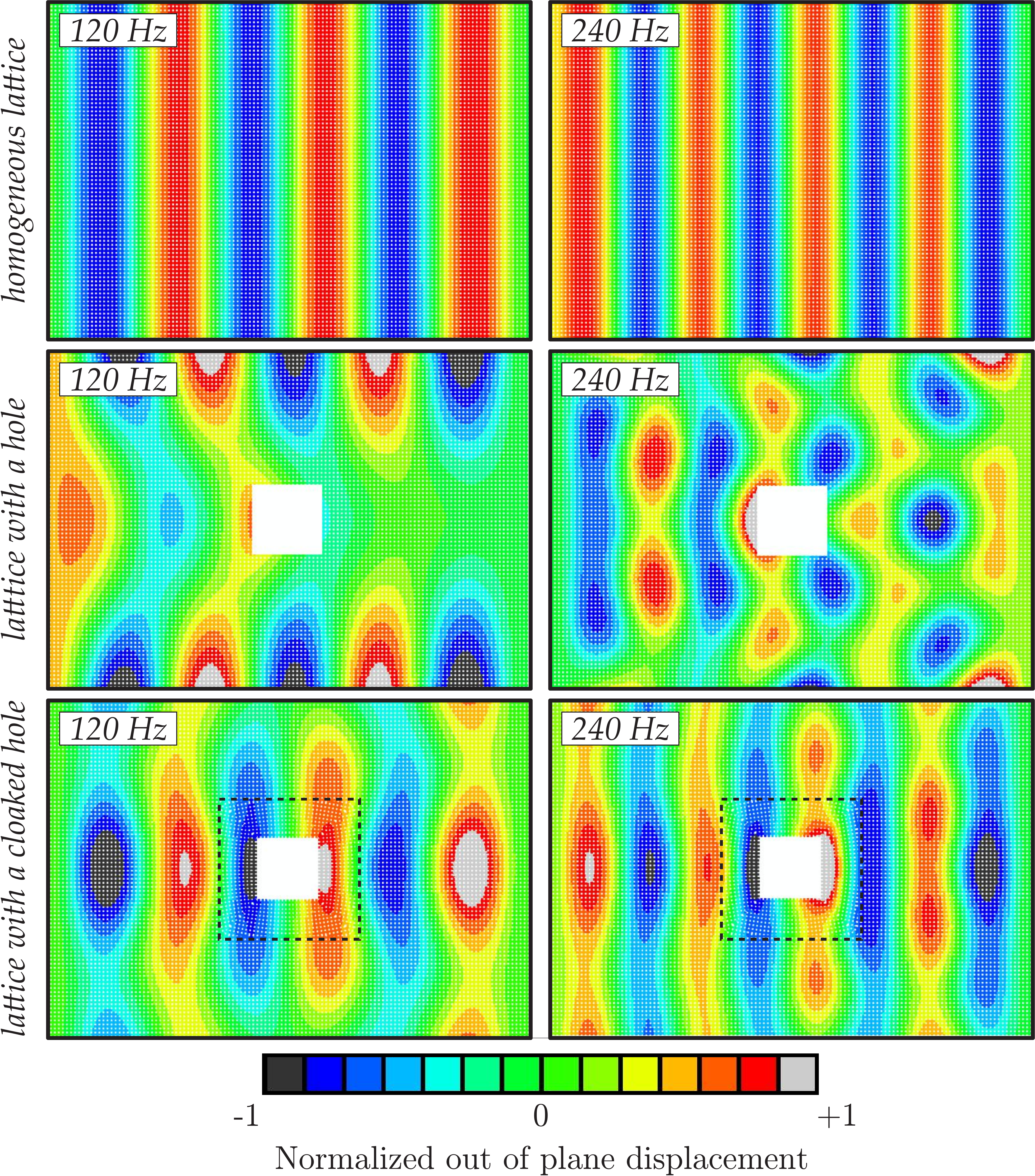}
\caption{\footnotesize Two-dimensional ABAQUS simulations performed in the case of a plate without a hole (upper), a plate with an uncloaked hole (middle), and a plate with a hole surrounded by the cloak (lower). For a frequency range between 120 Hz and 240 Hz,  the simulations show that the cloak reduces significantly the scattered field. }
\label{CloakEffect}
\end{center}
\end{figure}

In particular, the work by Milton, Briane, \& Willis~\cite{MiltonNJP,MiltonPRS} identifies the effects of `negative inertia' in the elastic cloak and analyses the requisite strong anisotropy, not only in the elastic compliance, but also in the inertial properties of the cloak.
The fact that, in the framework of Milton \emph{et al.}~\cite{MiltonNJP,MiltonPRS}, the mass density of the material should behave as a tensorial quantity rather than simply as a scalar is an important and striking observation.
A novel approach of dynamic homogenisation, capable of encapsulating such striking effects, was introduced and systematically studied by Craster and co-authors \cite{Craster1, Craster2}. The asymptotic theory leads to an effective equation for the envelope function in the perturbation approximation relative to a standing wave associated with a periodic lattice or continuum structure.
Norris \& Shuvalov~\cite{NorrisWaveMotion} later generalised the framework of Milton, Briane, \& Willis and showed that, although one cannot create an invisibility cloak for elastodynamics without recourse to a non-classical generalised theory of elasticity, with an appropriate choice of Gauge one can choose between creating a micropolar elastic cloak or a cloak with tensorial density, for example.
The concept, design, and theoretical analysis of elastodynamic invisibility cloaks are much more challenging compared to cloaks for membrane, acoustic, optical, and anti-plane shear waves, all of which are governed by the transformed Helmholtz equation.
Although the theoretical framework for elastic cloaks is well defined, its experimental implementation has never been successfully achieved for dynamic vector problems of elasticity.
The pioneering work by Wegener and his group for Kirchhoff-Love plates~\cite{Wegener2012} is the only significant experimental contribution in this extremely  challenging area.
In a different context, multi-scale resonators were discussed in \cite{Colombi} in relation to an approximate cloaking referred to as \lq Directional cloaking' for elastic plates containing voids.

The elastic Kirchhoff-Love plate provides an efficient and rigorous framework for the analysis of elastic waves in thin plates.
The propagation of flexural waves in a Kirchhoff-Love plate is governed by the biharmonic operator and waves in an homogeneous isotropic plate can be expressed as a linear combination of solutions of the Helmholtz equation (which we refer to as ``membrane waves'') and solutions of the modified Helmholtz equation.
The latter are evanescent fields but, nevertheless, may make a significant contribution through the boundary conditions and hence play a crucial role in the dispersion of flexural waves in structured plates~\cite{NatashaMovchan1,NatashaMovchan2,EvansPorter}, in addition to the design of flexural invisibility cloaks~\cite{JonesIntJSolsStruct}.

\begin{figure}
\renewcommand{\figurename}{\footnotesize{Fig.}}
  \begin{center}
      \includegraphics[width= 10 cm]{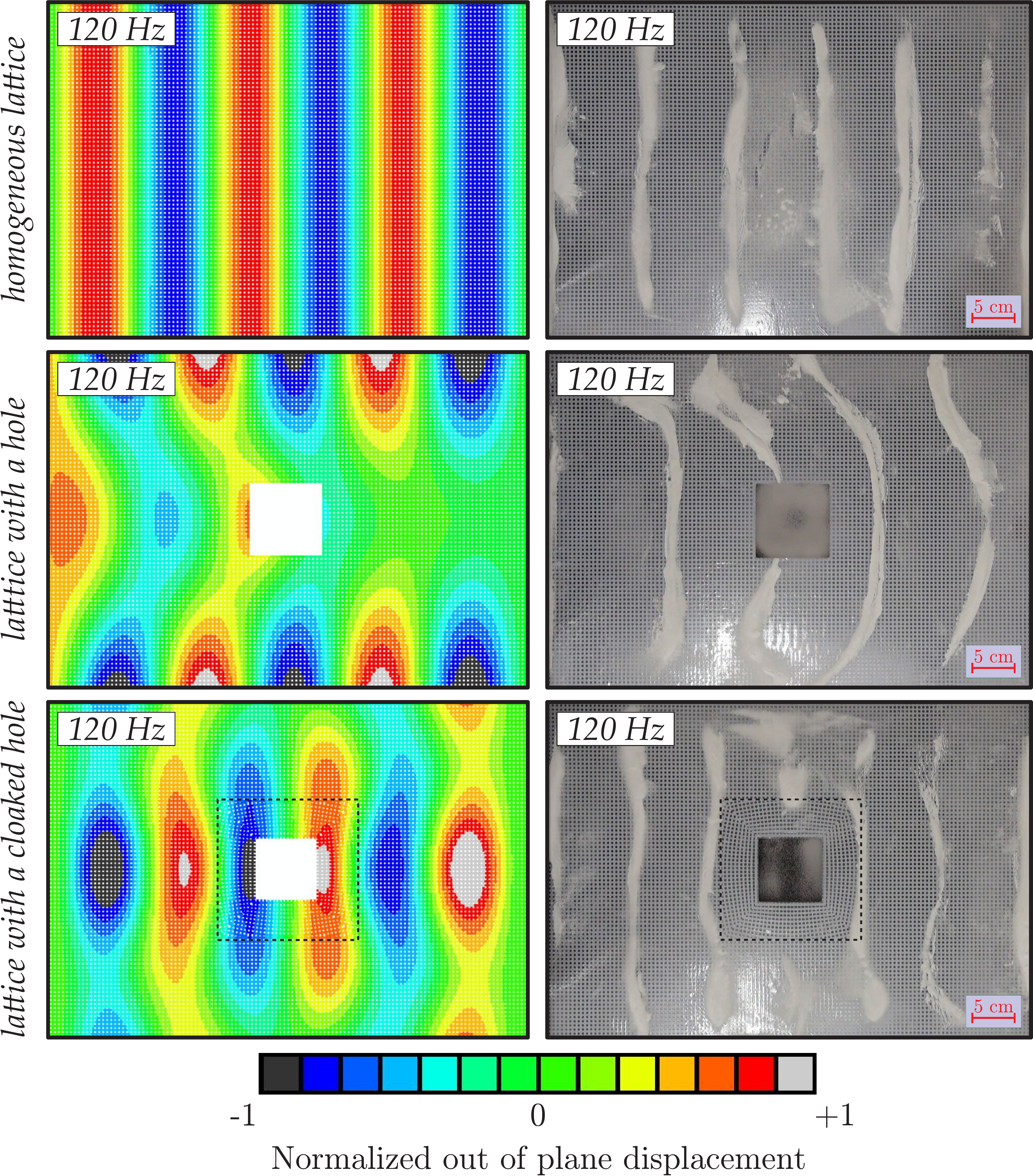}
\caption{\footnotesize { Comparison between experiments and numerical simulations in the case of an applied displacement with a frequency of 120 Hz}} \lb{Pict_Exp}
  \end{center}
\end{figure}

The design of the cloak presented here is distinct from the cloaks presented by Wegener \emph{et al.}~\cite{Wegener2012}, in so far as the cloak designed and constructed here is matched with a multi-scale mechanical structure, as opposed to a homogeneous continuous ambient matrix.
Moreover, the cloak which we construct here is not singular, i.e. it does not require infinite wave speeds on the interior boundary, but is based on the regularised cloaking transformation and theoretical design presented by Colquitt \emph{et al.}~\cite{ColquittPRS} for membrane waves and later for flexural waves in plates~\cite{ColquittJMPS}. 
The cloak design is shown in Fig.~\ref{Abaqus_geometry}A and the numerical and experimental implementations demonstrate that the invisibility cloak is efficient.
This is illustrated in Fig.~\ref{CloakEffect}, which shows a uniform vibrating plate without a hole, a plate with an uncloaked hole, and a plate with a hole surrounded by the cloak -- the latter reduces significantly the scattered field, as predicted by the analysis.

It is emphasised that here we consider the full elastodynamical problem of wave propagation in a discrete metamaterial lattice cloak embedded within a multi-scale ambient medium; this should be distinguished from the earlier work of Wegener \emph{et al.} on mechanical lattice cloaks~\cite{UnfealabilityCloak,WegenerPNAS} wherein the static cloaking problem is considered and no waves propagate within the system.

The structure of the paper is as follows. Section \ref{sqcloak} outlines the idea of the regularised cloaking transformation leading to the practical implementation of the multi-scale structured square cloak, which is surrounded by an ambient square lattice. 
The computational and experimental implementations are discussed in Section \ref{computational_exp}.
In particular, the computational section \ref{computational} presents several ABAQUS simulations to verify the efficacy of the cloaking effect, as shown in Fig. \ref{CloakEffect}.
The experimental design and results are discussed in section \ref{experiments}.
Finally, we draw together important concluding remarks in section \ref{concluding}.

\section{The Hooke-Chladni-Faraday visualisation}
\label{HCF}

Before presenting the mathematical model of the cloak and the experimental implementation, we first discuss an elegant technique that we will employ to visualise the time-harmonic vibration of plates.
The method, known as the \emph{``Hooke-Chladni-Faraday''} technique, has been employed by many researchers in physics and mechanics for almost four centuries and continues to be used today.
Indeed, the recent paper~\cite{Chen_Luo} examines the visualisation of standing waves in dynamically reconfigurable liquid-based templates, which were used to assemble micro-scale materials into ordered structures with desired properties.
This novel approach at a micro-scale level employed the idea of Faraday waves.
A three-dimensional visualisation of acoustic standing waves was reported in a recent paper \cite{3DStandingWave} that demonstrated the elegant efficiency of the classical Hooke-Chladni-Faraday method, which continues to generate new ideas and exciting results.
In particular, the experiment reported in  \cite{3DStandingWave}, shows levitating micro-particles along the nodal lines of a three-dimensional standing wave.

Well before the time of Michael Faraday, Robert Hooke and then Ernst Chladni discovered an ingenious method to visualise standing waves in elastodynamics.
This was especially effective for flexural resonances in elastic plates and membranes, as described in Chladni's book~\cite{Chladni}.
The technique consists of drawing a violin bow over a metallic or glass plate that is covered with flour.
Once the plate reaches a resonance the flour collects along the nodal lines of the resonant mode providing an elegant yet efficient visualisation of the standing wave present in the plate.

Following Chladni's experimental demonstration, the eigenvalue problem for the free vibrations of a square plate with free edges has been studied by many scientists, most notably, Lord Rayleigh~\cite{RayleighBook,Rayleigh1911} and Ritz~\cite{Ritz1908,Ritz1909} during the development of the now well known Rayleigh-Ritz method.
Although no closed form solution currently exists, Ritz~\cite{Ritz1909}, over a century ago, was able to construct remarkably accurate approximate solutions for the eigenfrequencies and nodal patterns.
These classical studies have led to a fascinating area of Cymatics, which analyses methods of making sound and vibration visible and has attracted attention of engineers, mathematicians, physicists and musicians around the world.

\begin{figure}[h]
\renewcommand{\figurename}{\footnotesize{Fig.}}
\begin{center}
\includegraphics[width=15 cm]{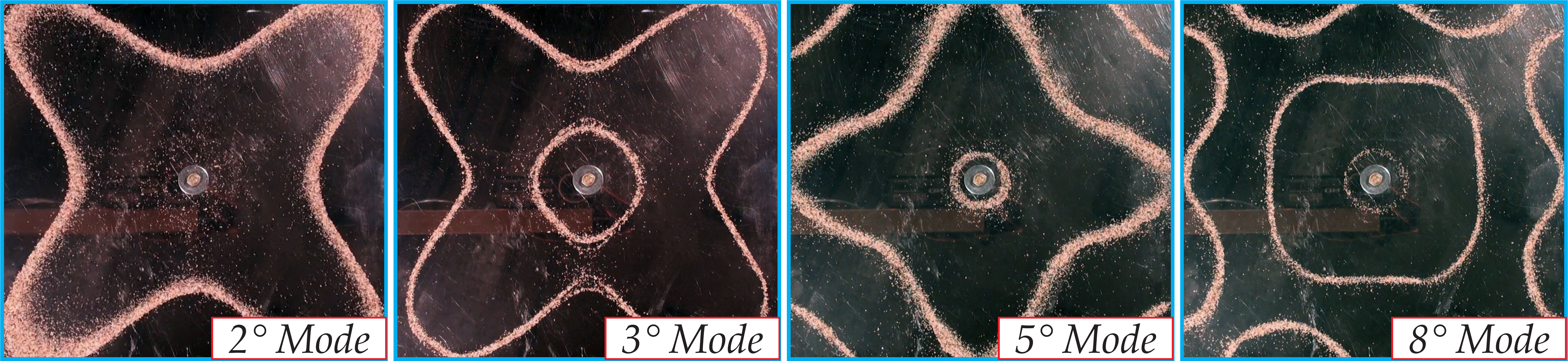}
\caption{\footnotesize Hooke-Chladni-Faraday visualization of four eigenmodes of a square elastic plate with a free boundary.
These illustrative experiments were produced in the \lq Instabilities Lab'  of the University of Trento. 
}
\label{Chladni_Figures}
\end{center}
\end{figure}

For the purpose of illustration, we include in Fig. \ref{Chladni_Figures}  examples of the Chladni patterns for eigenmodes of a square elastic plate with a free boundary, which are accurate and fully consistent with analytical findings. The patterns depend on the boundary conditions and on any inhomogeneities that may occur in the plate, such as voids or inclusions. In particular, Chladni patterns were never constructed for a plate with a hole surrounded by a structured cloak.

In the present paper, we visualise Hooke-Chladni-Faraday patterns for flexural waves around an obstacle surrounded by a multi-scale structured cloak and elegantly illustrate the efficacy of the cloak. An ABAQUS simulation for the mechanical configuration, identical to the one used in the experiment, provides accurate numerical data on the displacement amplitudes and stress distribution.
We present the experimental visualisation to confirm the predicted wavefront profiles.

Compared to the classical settings for the above mentioned frequency response problem for a rectangular Kirchhoff plate, we go further and consider a structured plate with a cloaked hole.
Fig. \ref{Pict_Exp} includes Hooke-Chladni-Faraday patterns used for visualisation of the cloaking effect of flexural waves. These observations are new and demonstrate scattering patterns for three configurations, which include (a) a rectangular lattice-type plate, (b) a plate with a square hole, and (c) a plate with a structured cloak enclosing the hole.
As in the original experiments by Hooke, the powder collects along the nodal lines of the vibrating plate thus indicating the position of the wavefronts, i.e. the locus of points on the wave with the same phase and zero displacement.
The Hooke-Chladni-Faraday patterns allow us to conveniently visualise the propagation of waves in the structured plate and, in particular, we can observe a significant reduction of the scattering pattern for the case (c) (as in the bottom part of Fig. \ref{Pict_Exp}). A detailed discussion of the experiment is given in the main text of the paper.

\section{The square cloak}
\label{sqcloak}

In most of the papers addressing the design of cloaks for linear waves, a singular radially symmetric  push-out transformation is employed (see, for example, \cite{Pendry, ParnellPRS}) which maps a point to a finite disc.
For theoretical and computational models of continuous media, such an approach is adequate.
However, for practical implementations it poses substantial difficulties.
In particular, not only does the transformation lead to infinite wave speeds on the interior boundary of the cloak, but also the required cloaking material is characterised by an unrealistic strong anisotropy.
Indeed a lattice structure, instead of a continuum, can be used to create an invisibility cloak.
A significant advantage of lattices is that they naturally accommodate high contrasts in their compliance leading to strong anisotropy.
However, a radial discrete cloak that fits inside a circular ring would not match any periodic lattice, and the presence of a geometrical mis-match on the interface boundary leads to a substantial mis-match in the interface boundary conditions.

For the design presented here, we choose to employ a square cloak following the theoretical framework recently established in~\cite{ColquittJMPS}.
The cloak is embedded in an ambient square lattice, which is subjected to out-of-plane flexural vibrations.
We consider a thin structured plate with a defect represented by a traction free void, which significantly influences the wave field, as illustrated in Fig.
\ref{CloakEffect}.
A structured cloak, as in \cite{ColquittPRS,ColquittJMPS}, is then installed around the void.
In the cloaked configuration, we observe a significant reduction in the scattered field and, in particular, the reduction in the shadow region behind the scatterer and the restoration of the incident field represented by a plane wave.

\subsection{The regularised cloaking transformation}

\begin{figure}
\renewcommand{\figurename}{\footnotesize{Fig.}}
\begin{center}
\includegraphics[width=15.0 cm]{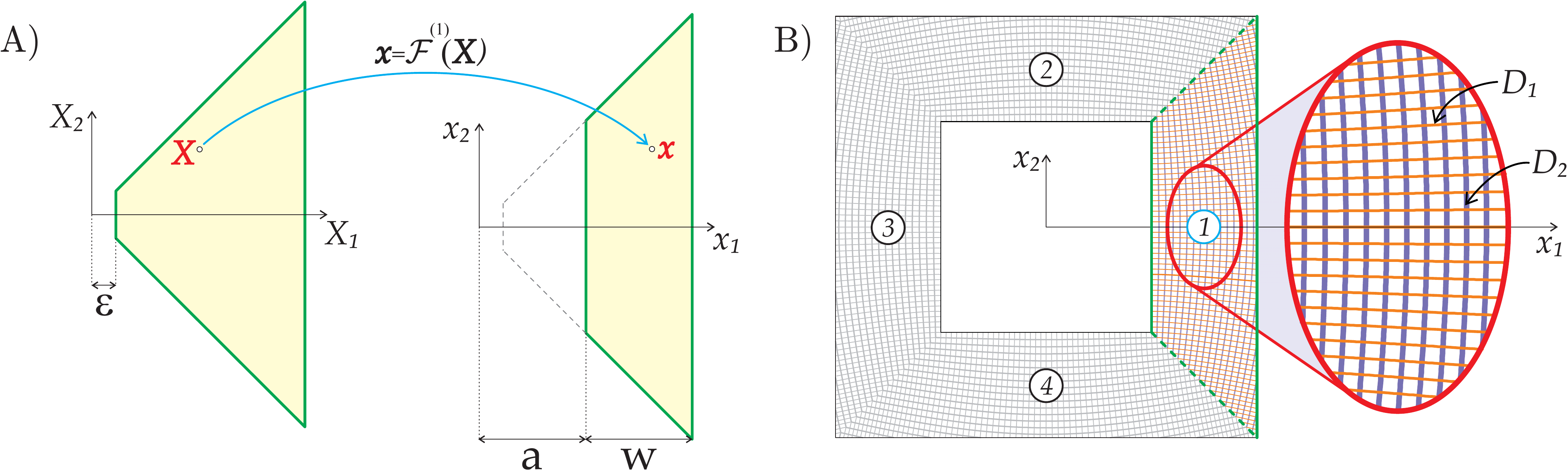}
\caption{\footnotesize (A) The function $\vec{\mathcal{F}}^{(1)}$ maps the undeformed trapezoidal region (1)  to the deformed trapezoidal configuration. (B) A discrete lattice structure where the curved ligaments are oriented following the principal directions of the stiffness matrix for the continuum cloak.}
\label{Deformation_Schematic}
\end{center}
\end{figure}

Following Colquitt et al. \cite{ColquittJMPS}, we choose a regularised near-cloak, which is obtained by four push-out transformations applied to trapezoidal regions, as illustrated in Fig. \ref{Deformation_Schematic}A.
The idea of regularisation is to map a domain with a small hole (e.g. a square of semi-width $\epsilon$ as in Fig. \ref{Deformation_Schematic}A) at the centre into another domain, whose exterior boundary is preserved while the interior boundary is expanded to the required finite size.
For the regularised problem, we set the Neumann boundary condition on the boundary of the hole, which means (in the case of an elastic plate) the free edge boundary condition.
As showed recently~\cite{JonesIntJSolsStruct}, the correct choice of boundary condition is of vital importance in order to achieve cloaking.

The mapping is defined in such a way  that ${\bf x} = \vec{\mathcal{F}}^{(i)}({\bf X})$ for each trapezoidal region ($j=1,2,3,4$) shown in Fig.\ref{Deformation_Schematic}A. In particular, the map for the trapezoidal region (1), with bases perpendicular to the $X_1-$axis,  is defined by the formula:
\beq
\vec{\mathcal{F}}^{(1)}({\bf X}) = \begin{bmatrix}
\alpha_1 X_1 + \alpha_2 \\
\alpha_1 X_2 + \alpha_2X_2/X_1
\end{bmatrix},\qquad
\label{FF}
\eeq
where
\beq
\alpha_1=  \frac{w}{a+w-\epsilon},  \hspace{1cm} \alpha_2=  \frac{(a+w)(a-\epsilon)}{a+w-\epsilon}.
\label{coeff}
\eeq
Here, $0<\epsilon\ll1$ is the regularisation parameter and uppercase letters denote the undeformed configuration, whilst lowercase letters denote the deformed configuration.  Prior to the transformation, the interior boundary of the trapezoidal region (1) corresponds to $X_1=\epsilon$. Following the transformation, the interior boundary of this  trapezoidal region is moved to $x_1=a$.
The outer boundary of the cloak is invariant with respect to the transformation, and for the region (1), it corresponds to $X_1 = a+w$.
The other three trapezoidal regions are transformed in the same way, subject to a rotation.

\subsection{Transformed equations after the cloaking transformation}

The time-harmonic flexural deformations of a homogeneous thin elastic plate are governed by the Kirchhoff plate equation, which before the transformation has the form
$$
\Big(\nabla_X^4 - \frac{\rho h}{D} \omega^2 \Big) w({\bf X}) =0.
$$
After the application of a non-singular transformation we have
\beq
\Big(\nabla \cdot J^{-1} {\bf F} {\bf F}^{T} \nabla J \nabla \cdot J^{-1} {\bf F} {\bf F}^{T} \nabla  - \frac{\rho h}{J D} \omega^2 \Big) w({\bf x}) =0,
\label{transeq}
\eeq
where $ {\bf F} = \nabla_X {\bf x}$ is the non-degenerate Jacobi matrix of the geometrical transformation, and $J = \det {\bf F}.$
The above equation can be interpreted as the equation of an anisotropic pre-stressed plate, as discussed in \cite{ColquittJMPS}.
In the context of asymptotic approximations, the paper \cite{BrunNJP} discusses the configurations where the effects of pre-stress are small, and when an approximate factorisation of the transformed differential operator is possible.

Our goal is to construct an approximate cloak, which may be conveniently implemented experimentaly.
We observe that in the successful experiment for the continuous plate led by Wegener \cite{Wegener2012}, the authors assumed that the membrane waves are dominant and configured their radially symmetric approximate cloak accordingly.
Although such an assumption may appear to be inappropriate, it has been demonstrated that, within a certain frequency range, the membrane waves indeed are dominant and the approximate cloaking effect is apparent.
This approach is further reinforced for the square cloak by the observation made in \cite{ColquittJMPS} that the principal directions of stiffness for the membrane cloak and the flexural cloak are exactly the same.
Compared to implementations of invisibility cloaks for continua, such as in~\cite{Wegener2012}, the square lattice cloak developed in the present paper is relatively straightforward.
We now proceed to describe the implementation of our novel square lattice cloak.

\subsection{The lattice approximation of the cloak}

\begin{figure}
\renewcommand{\figurename}{\footnotesize{Fig.}}
\begin{center}
\includegraphics[width=13.0 cm]{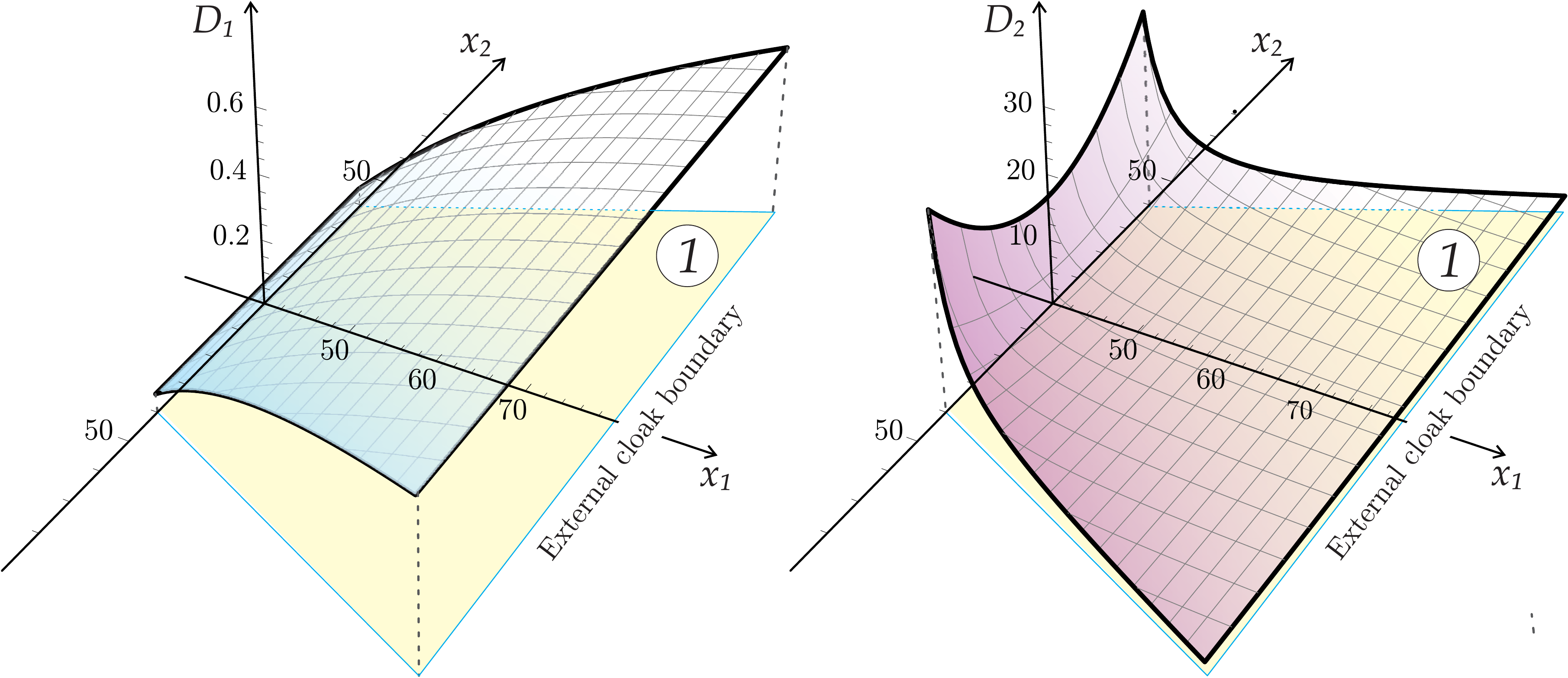}
\caption{\footnotesize The required stiffnesses $D_1$ and $D_2$ for the cloak ligaments reported as function of $x_1$ and $x_2$ .}
\label{Rigidity}
\end{center}
\end{figure}

We consider an approximate cloak, realised using a discrete lattice structure with curved ligaments as derived in \cite{ColquittJMPS}.
These elastic ligaments are aligned with the the principal directions of the stiffness matrix for the continuum cloak, as
illustrated in Fig. \ref{Deformation_Schematic}B, and yield a structured flexural lattice system.
The thin elastic ligaments may be treated as beams of rectangular cross section characterised by the bending stiffnesses $D_1$ and $D_2$ chosen in accordance with the analytical formulae from \cite{ColquittJMPS}
\beq
D_1=  \frac{x_1-\alpha_2}{x_1},   \hspace{1cm} D_2=  \frac{x_1^4+\alpha_2^2 x_2^2}{x_1^3(x_1-\alpha_2)}.
\label{rigidcoeff}
\eeq
These flexural rigidities are visualised in Fig. \ref{Rigidity} for the right-hand quadrant.
For this quadrant, on the $X_1$ axis,  $D_1$ represents the $X_1$ stiffness in the principal direction, whereas $D_2$ represents the $X_2$ stiffness in the other principal direction.
In every other point inside the cloak, $D_1$ and $D_2$ represent the stiffnesses in the principal directions of the locally orthotropic cloak constructed here.
We note that on the interior boundary of the cloak the tangential rigidity $D_2$  is much higher compared to the normal rigidity $D_1$ 
and emphasise that these stiffnesses are finite on this boundary of the cloak (see Fig. \ref{Rigidity}).

According to equation (\ref{transeq}) the mass density $\rho'$ inside the cloak is also non-uniform and obeys the formula
\beq
\rho' = \rho \frac{x_1-\alpha_2}{\alpha_1^2 x_1},
\label{density}
\eeq
where $\rho$ is the density outside the cloak and the coefficients $\alpha_1$ and $\alpha_2$ are defined in (\ref{coeff}).

Equations (\ref{rigidcoeff}) and (\ref{density}), together with the accompanying geometrical design of the cloak, provide the essential information for the experimental implementation discussed in the text below.

\section{Numerical simulation and experimental verification}
\label{computational_exp}

We have performed both finite element (FE) computations, in ABAQUS, and also experiments to verify the efficacy  of the square structured cloak in reducing the scattering of flexural waves from voids.
The structured cloak has been designed and implemented in SOLIDWORKS.
Each elastic ligament has a specified variable cross section that provides the required rigidities $D_1$ and $D_2$ as reported in Colquitt \emph{et al.} \cite{ColquittJMPS}.
The physical lattice cloak was created by milling holes into a polycarbonate plate;
both ABAQUS and the milling machine were programmed using the same SOLIDWORKS  original code. In so doing, experiments and simulations have identical geometry, material parameters, constraints and applied out-of-plane vibrations.
Figure \ref{Abaqus_geometry} illustrates both the lattice geometries implemented in ABAQUS and the experiment.

\subsection{ABAQUS simulation}

\label{computational}

We have compared the wave field for three cases: the first for  a homogeneous lattice in the absence of any void, the second in the presence of a void, and the third in the presence of a void surrounded by our specially designed invisibility cloak.
The simulations have been performed using a parametric python script for ABAQUS, 
run by means of MATLAB.
We have computed  the steady-state frequency response of each lattice using the Dynamic/Explicit package already implemented in ABAQUS.
Since the elastic lattices are constructed from thin elastic ligaments, we have  performed the simulations employing 3D beam elements. 
\begin{figure}
\renewcommand{\figurename}{\footnotesize{Fig.}}
  \begin{center}
      \includegraphics[width= 14 cm]{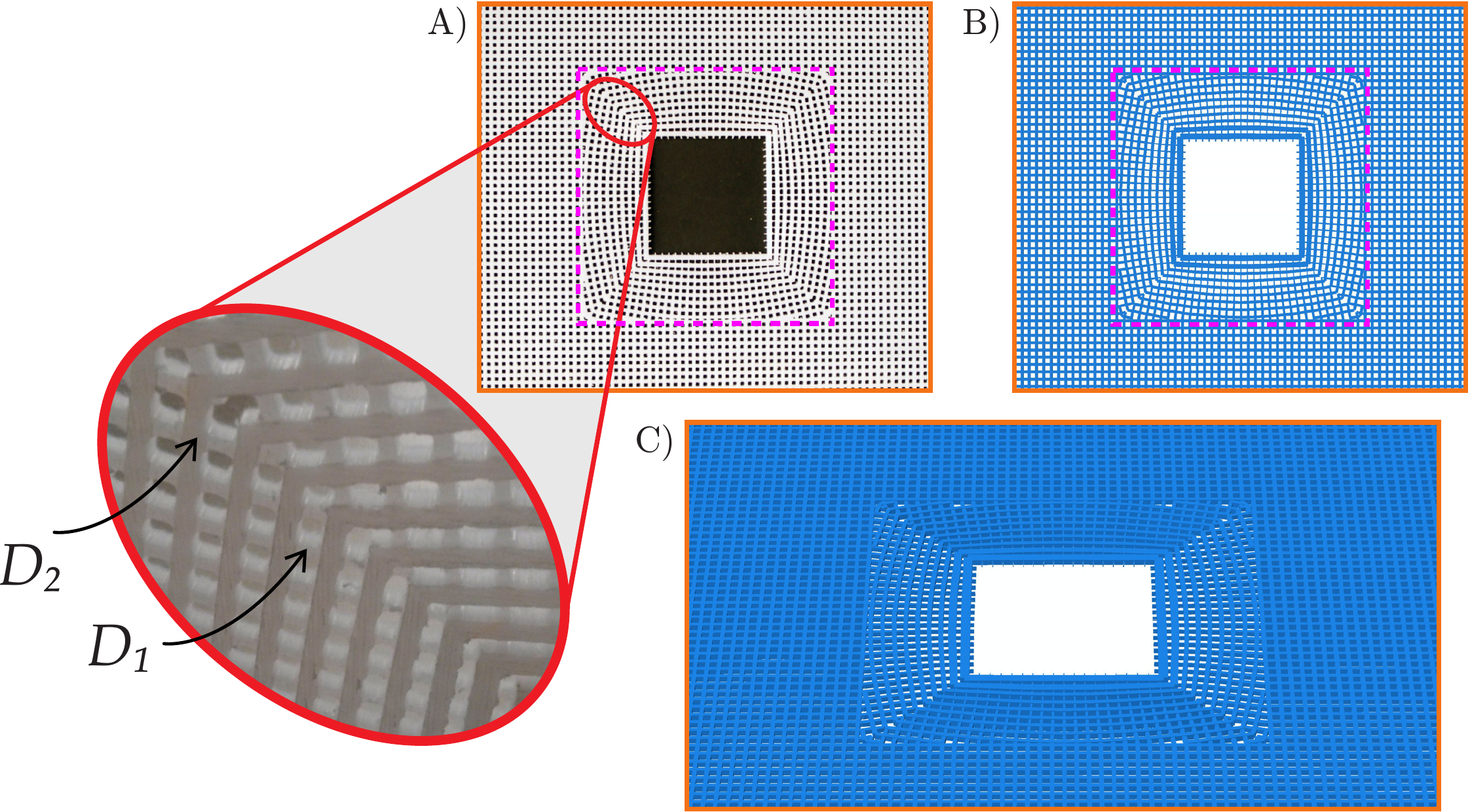}
\caption{\footnotesize {The comparison between the geometry of the real cloaked lattice drilled by a milling machine (A) and a frontal view (B) and a 3D view (C) of the geometry implemented in ABAQUS.
The zoomed area of the part (A) shows {the physical} realisation of the square cloak embedded in a square lattice. By observing the enlarged detail of the square cloak, the different cross sectional areas of the ligaments are clearly visible and chosen to match the required $D_1$ and $D_2$ principal flexural rigidities.
}} \lb{Abaqus_geometry}
  \end{center}
\end{figure}

The cloak design is an approximate one, in the sense that it works well for a certain range of frequencies. 
Using the simulations we were be able to determine the range of frequencies over which the cloak is effective.
An out-of-plane vibration in the range between 100 Hz and 250 Hz has shown the predicted cloaking action.  In particular, in Fig. \ref{CloakEffect} we show the cases of 120 and 240Hz, where the presence of the cloak leads the holed plate to behave in a similar way to an homogenous lattice without a void.

\subsection{The experimental implementation}
\label{experiments}

The model,  designed theoretically and implemented  in ABAQUS, has also been verified experimentally (the geometry is as in  Fig. \ref{Abaqus_geometry}).
The experiment has been performed at the \lq Instabilities Lab' of the University of Trento.
In the current setup, we use the Hooke-Chladni-Faraday visualisation described earlier in Section \ref{HCF}.

Three structured flexural systems, namely a homogenous lattice, a lattice with a hole, and a lattice with a cloaked hole, have been produced  by drilling polycarbonate plates (white 2099 Makrolon UV from Bayer) with an EGX-600 Engraving Machine (accuracy 0.01mm, by Roland).
The mechanical properties of the polycarbonate, namely  elastic modulus (E), density ($\rho$) and Poisson's ratio ($\nu$), are respectively 2350 MPa, 1200 kg/m$^3$ and 0.35.
The lattice cloak as well as the geometrical and material parameters are chosen consistently, according to the same data set stored in the
SOLIDWORKS file, as illustrated in Fig. \ref{Abaqus_geometry}A, which shows the experimental sample.
Thin transparent film was used to cover the flexural lattice system to enable the use of powder for the Hooke-Chladni-Faraday visualisation.

The lattices, externally  measuring 600mm by 400mm, have ligaments  of constant cross-section (1.75 mm by 2.5mm) outside the cloak, whereas within the cloak the ligaments have variable width and height according to the analytical algorithm of Section \ref{sqcloak}.

The lattices, constrained by clamps on the two shorter sides and having the other two free, have been excited by using a TIRA Vibrations Test System TV51144 and BOSE ElectroForce 3300 Series II, connected to the left clamp of the lattices, as shown in the experimental arrangement Fig. \ref{Exp_Setup}.

The maximum amplitude (1 mm peak to peak) of the sinusoidal displacement has been imposed by the oscillating clamp connecting the Vibrations Test System to an NI CompactRIO system, interfaced with LabVIEW 2014 (National Instruments).

\subsection{The experimental results and comparison with the computational model} 

The qualitative assessment of the effect of the cloak has been carried out using the Hooke-Chladni-Faraday technique  that shows the positions of the nodal lines of the vibrating plate.

The boundary conditions are chosen consistently, both in the numerical simulation, and the physical experiment: one side of the plate is rigidly clamped, while the opposite side is attached to a moving clamp and excited by applying a time-harmonic displacement. The remaining two sides are traction free, i.e. the bending moments and the transverse forces are equal to zero at these boundaries.

\begin{figure}
\renewcommand{\figurename}{\footnotesize{Fig.}}
  \begin{center}
      \includegraphics[width= 14 cm]{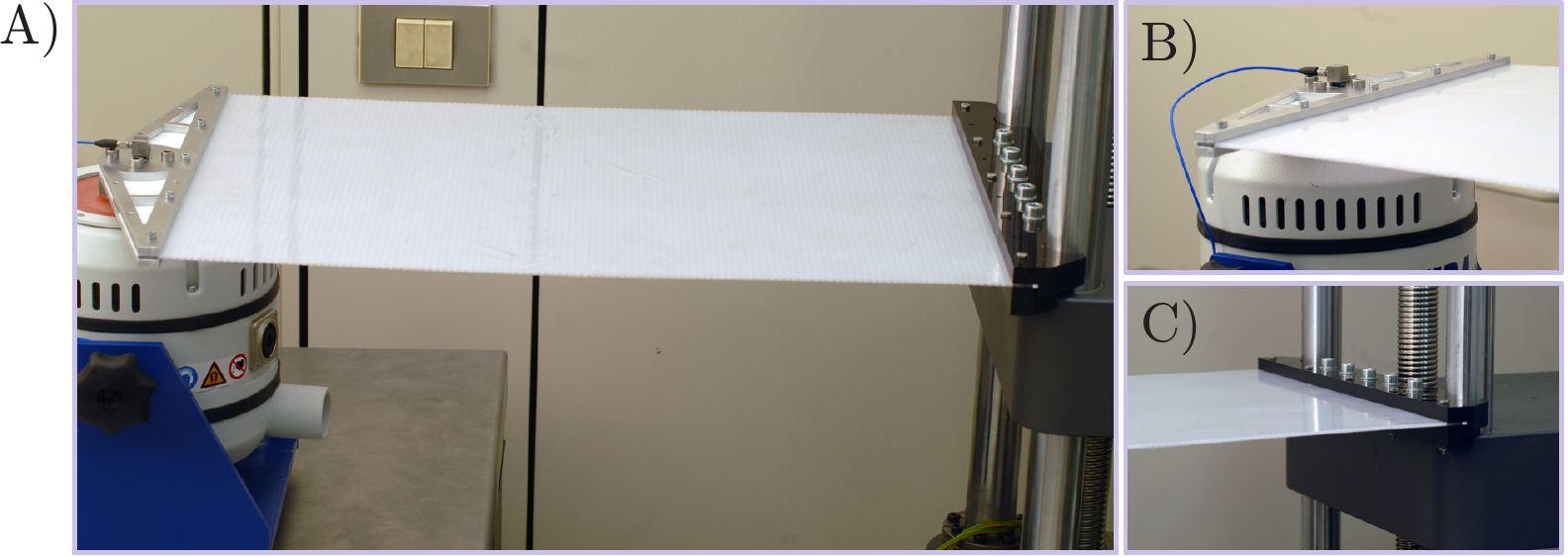}
\caption{\footnotesize {The vibration apparatus employed in the
experiments (A) the details of the constraints on both sides of the plate, namely a shaking clamp (B) and a rigid clamped (C).}} \lb{Exp_Setup}
  \end{center}
\end{figure}

Photos have been taken with a Sony NEX 5N digital camera (equipped with 3.5-5.6/18-55 lens, optical
steady shot manufactured by Sony Corporation) and with a Nikon D200 digital camera (equipped with a AF-S micro Nikkor lens 105 mm 1:2.8G ED).

Fig. \ref{Pict_Exp} shows the comparison between the experiment and the numerical simulations in the case of an applied displacement with a frequency of 120 Hz.

In particular, in the upper part of Fig. \ref{Pict_Exp}  we see the unperturbed plate without a hole, with the one-dimensional frequency response pattern shown. 
In the middle part of Fig.   \ref{Pict_Exp} we show that plate containing an uncloaked square hole, with
a traction free boundary.

We observe similar shadow regions both, in the ABAQUS simulation, and the physical experiment with a real structured plate.

Finally, we demonstrate in the lower part of Fig.  \ref{Pict_Exp} that the shadow region is significantly reduced when the cloaking region, depicted in Fig. \ref{Abaqus_geometry}A, is introduced around the hole, so that the wave pattern outside the hole becomes almost flat, as expected.
It is clear that there is good qualitative agreement between the experimental and numerical results.
In fact, in the case of a cloaked void, the nodal lines showing the incident field represented by a plane wave are almost straight, similar to the case of the homogeneous lattice. On the other hand, in the absence of the cloak, the pattern lines appear deeply influenced by the hole.
In this case we observe rounded nodal lines that differ significantly from the straight wavefronts observed for the homogenous lattice.

\section{Concluding remarks}
\label{concluding}

In this paper, we have presented a proof of concept design for a square invisibility cloak.
Having constructed the cloak, we proceeded to examined its effectiveness using both, in a computational ABAQUS model, as well as in real physical experiment.
This novel design, proposed in an earlier theoretical paper \cite{ColquittJMPS}, was appealing due its simplicity and elegance, which made an experimental implementation feasible.
The regularisation introduced into this design of cloak enables careful and precise implementation of boundary conditions on the interior boundary of the cloak in both the numerical simulations and the experiments.

The approximate cloak presented here, proves to be efficient within a predicted frequency range, but the results become frequency sensitive as the frequency of the incident wave increases.
This effect has been expected, and similar phenomena of the high frequency sensitivity were noted in \cite{Wegener2012}.

The range of applications of the proposed cloaking device is wide and it covers, in particular, earthquake resistant systems, as well as novel designs of foundations of civil engineering structures.

\vspace*{5mm} \noindent
{\sl Acknowledgments } The authors would like to thank Prof Davide Bigoni for stimulating discussions and valuable suggestions on the text of the paper, and for his strong support. D.M. and A.B.M. gratefully acknowledge financial support from the ERC Advanced Grant \lq Instabilities and nonlocal multiscale modelling of materials' FP7-PEOPLE-IDEAS-ERC-2013-AdG (2014-2019) and provision of experimental facilities via the \lq Instabilities Lab' of the University of Trento (http://www.ing.unitn.it/dims/ssmg/).  N.V.M gratefully acknowledge financial support from European FP7 - INTERCER-2 project (PIAP-GA-2011-286110-INTERCER2).
D.J.C. thanks the EPSRC for support through research grant EP/J009636/1.

\end{document}